\PassOptionsToPackage{dvipdfmx}{graphicx}
\documentclass[preprint,pteplogo]{ptephy_v2}

\preprintnumber{2511-028} 
\usepackage{hyperref}
\usepackage{ulem}
\usepackage{subcaption}
\usepackage{bm}

\newcommand{\Const}{\mathrm}
\newcommand{\Var}{\mathit}

\begin{document}

\title{Wave Front Sensing demodulated at the difference frequency between two phase-modulation sidebands in a compound interferometer configuration for a gravitational-wave detector}


\author{Chiaki Hirose}
\affil{Graduate School of Science and Technology, Niigata University, Nishi-ku, Niigata
City, Niigata 950-2181, Japan \email{f22l003c@mail.cc.niigata-u.ac.jp}}

\author{Kenta Tanaka}
\author{Osamu Miyakawa} 
\author{Takafumi Ushiba}
\affil[2]{Institute for Cosmic Ray Research (ICRR), KAGRA Observatory, The University of
Tokyo, Kamioka-cho, Hida City, Gifu 506-1205, Japan}



\begin{abstract}
   Precise alignment sensing and control are essential for maintaining the stability of laser interferometric gravitational-wave detectors. Conventional Wave Front Sensing technique (WFS), which relies on the beat between the carrier and phase-modulated (PM) sidebands, is dominated by arm-axis signals when the carrier resonates in the full interferometer. This dominance limits the detection of other optical axes, such as the Power Recycling Cavity (PRC) and incident beam axes. To address this problem, we propose a novel sensing technique, “Phase-Modulated-sideband $\times$ Phase-Modulated-sideband Wave Front Sensing” (PMPMWFS), which demodulates the beat signal at the difference frequency between two anti-resonant PM sidebands. We derived the theoretical response of PMPMWFS and experimentally demonstrated it using the Power-Recycled X-arm (PRXARM) configuration of KAGRA. The results show that PMPMWFS effectively decouples angular fluctuation signals of the PRC and incident beam from those of the arm cavity and provides orthogonal signal components for the end mirror of the arm cavity. Furthermore, feedback control using PMPMWFS achieved stable interferometer locking for over one hour. These results demonstrate that PMPMWFS offers an effective sensing method for decoupling multiple alignment degrees of freedom in future gravitational-wave detectors.
\end{abstract}

\subjectindex{XXXX, XXX}

\maketitle

\section{Introduction.}
Gravitational Waves (GWs) are ripples in spacetime that propagate at the speed of light, generated by the accelerated motion of massive objects \cite{Annalen_der_PhysikVolume_49}. Observable GWs are emitted by astrophysical phenomena such as the mergers of compact binaries, the rotational motion of pulsars, and supernova explosions. 
The direct detection of these GWs is expected to yield significant scientific outcomes, including measurements of the Hubble constant \cite{Soares-Santos_2019} and the verification of general relativity \cite{PhysRevD.103.122002}. This is achieved using laser interferometric GW detectors, each of which combines multiple optical resonators.\par

Determining the direction of the GW source using laser interferometric GW detectors requires the detection of the time difference in the arrival of the GW at multiple detectors. As an example, on 14 August 2017, simultaneous GW observations were achieved for the first time by Advanced LIGO \cite{Aasi_2015} and Advanced Virgo \cite{Acernese_2015}, with greater precision than in previous observations \cite{PhysRevLett.119.141101}. Therefore, it is important to operate GW detectors stably over the long term.\par

In laser interferometric GW detectors that incorporate multiple optical cavities, angular fluctuations of the mirrors constituting the interferometer can cause a relative misalignment between the incident beam axis and the resonant cavity axes, potentially reducing the stability of the interferometer. The interferometer stability can be improved by detecting this relative misalignment and feeding it back to the mirrors. \par
Wave Front Sensing technique (WFS) is primarily employed to detect optical axis misalignment in interferometers \cite{Morrison:94}. The laser beam incident on the interferometer with multiple optical resonators is phase-modulated to generate sidebands using an Electro Optic Modulator (EOM). The beam reflected from the resonator is detected by a Quadrant Photo Diode (QPD). By demodulating the detection signal at the phase modulation frequency, the relative misalignment between the optical axis of the phase-modulated sideband (PM sideband) and that of the carrier is measured.\par
Japanese GW detector KAGRA \cite{10.1093/ptep/ptac166} conducted joint observations with LIGO during the fourth international collaborative GW observation run (O4a) \cite{Ushiba:2024mR}, from 24 May to 20 June 2023 (UTC). KAGRA employed a Power Recycled Fabry-Pérot Michelson Interferometer (PRFPMI) during O4a. The PRFPMI consists of a Fabry-Pérot Cavity (FPC) in each arm of the Michelson Interferometer (MI) called X- and Y- arm cavity, and the Power Recycling Mirror (PRM) recycles the light reflected from the interferometer. The Power Recycling Cavity (PRC) increases the intracavity power in the FPCs at both arms, reducing the shot noise that limits the sensitivity. The primary degrees of freedom of the optical axis in the PRFPMI are the FPC axes of both arms, the short MI optical axis from the front mirrors of the FPCs to the BS, the PRC axis, and the incident light axis. 
Increasing the beam size reduces both the thermal noise of the mirror and that of the coating.\cite{PhysRevD.57.659} Therefore, KAGRA employs a large beam size for the FPC mirrors. However, this makes the optical axis more prone to tilting than in the other cavities. 
This FPC configuration is not unique to KAGRA but is also common to other large-scale laser interferometric GW detectors, such as LIGO and Virgo.
\par

When the carrier component resonates in the entire interferometer including the arm cavities, the WFS signal using the carrier detects changes in all axes.
Consequently, the signal from the arm cavity makes it difficult to discern the signals from other axes in the WFS signal, because the arm cavity axis is the most sensitive to the angular motion of mirrors. 
As a result, it becomes difficult to stabilize the interferometer by feedback control. In the future, KAGRA will be upgraded to the Resonant Sideband Extraction (RSE) \cite{MIZUNO1993273} to broaden the detection bandwidth for GW signals. Due to the increased number of degrees of freedom requiring control, this issue will become more pronounced, making signal separation crucial.\par
\begin{figure}[!h]
\centering
\includegraphics[width=6.0in]{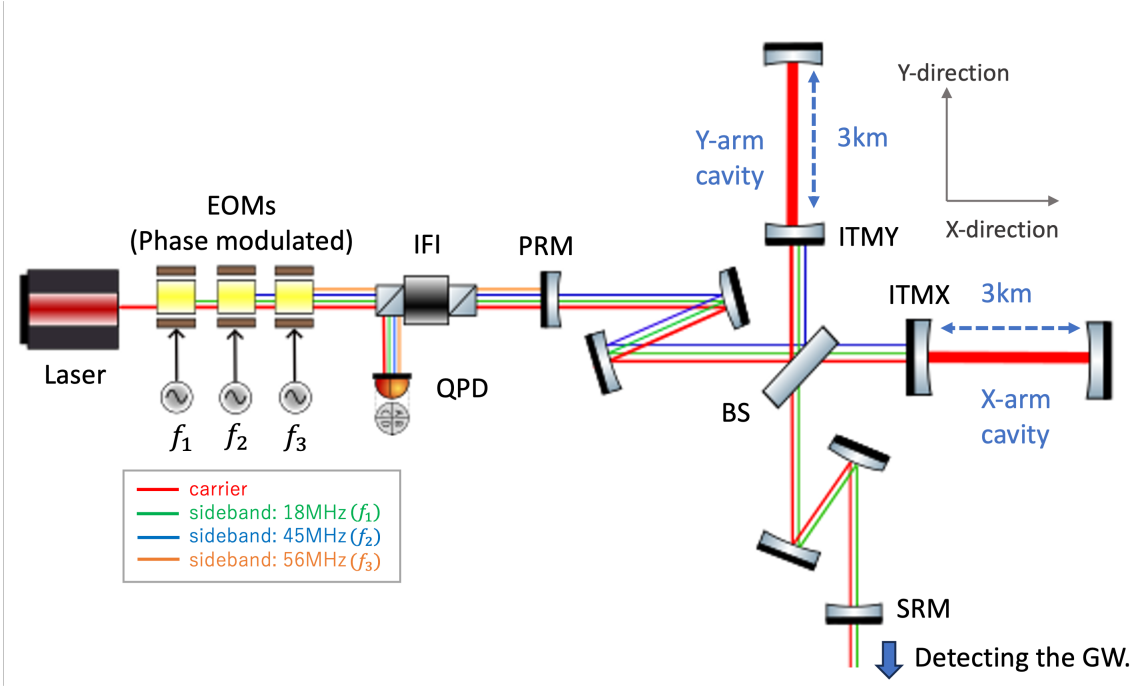}
\caption{
   RSE schematic diagram in KAGRA. The incident light is split by the BS into the X- and Y-directions. FPCs are installed in both arms of the MI, referred to as the X- and Y-arm cavity. The PRM forms the PRC with the two input mirrors of the FPCs (ITMX and ITMY). Similarly, the Signal Recycling Mirror (SRM) forms the Signal Recycling Cavity (SRC) with the ITMs. The gravitational wave signal is detected at the \mbox{SRC's} output port as a change in the optical path difference between the two arm cavities. Three PM sidebands are generated by the EOMs. The carrier beam resonates in both the PRC and the arm cavities. The PM sideband fields at $f_\Const{1}$, $f_\Const{2}$, and $f_\Const{3}$ exhibit distinct resonance states: the sideband field at $f_\Const{1}$ resonates with both the PRC and SRC, the sideband field at $f_\Const{2}$ resonates only with the PRC, and the sideband field at $f_\Const{3}$ is anti-resonant with all resonators. The reflected light from the interferometer is redirected by the Input Faraday Isolator (IFI) and detected by the QPDs.
   }
\label{FIG:RSE_KAGRA}
\end{figure}

In this paper, we propose a new alignment detection scheme, “Phase-Modulated-sideband $\times$ Phase-Modulated-sideband Wave Front Sensing” (PMPMWFS), which uses the difference frequency of two PM sidebands. This method is designed to detect the optical axis difference between the cavity axis and the incident axis to the cavity by tuning PM sideband frequencies so that one of them resonates on the cavity and the other does not. Specifically, the two sidebands adopted here (Fig. \ref{FIG:RSE_KAGRA}) do not resonate in the arm cavities. Therefore, the ratio of the arm cavity axis fluctuation signal to the signals from other optical axes detected by the PMPMWFS is expected to be smaller than that in the WFS signal. If the PMPMWFS can achieve signal separation between the fluctuations of the incident beam and the PRC axis from those of the arm cavity, it will provide an effective means of alignment control to stabilize the interferometer.\par

This paper summarizes the derivation of the PMPMWFS signal in Sec. \ref{SEC:PMPMWFS_CAL}. In Section \ref{sec:PRXARM}, we present the results of measuring the PMPMWFS signal in the PRXARM, in which \mbox{KAGRA's} PRC and X-arm cavity are resonant, and demonstrate alignment control of \mbox{KAGRA's} interferometer using the PMPMWFS signal.\par

\section {Derivation of the PMPMWFS.}
\label{SEC:PMPMWFS_CAL}
This section presents the theoretical derivation of the PMPMWFS signal for a composite optical resonator. Section \ref{SUBSEC:cal_model} describes the interferometer configuration of the composite optical resonator used in the theoretical analysis of the PMPMWFS signal. Section \ref{SUBSEC:refl_electricfield} shows the electric field of the reflected light from the composite optical resonator for the fundamental Gaussian mode and the first-order Hermite-Gaussian mode. Section \ref{SUBSEC:demosignal} details the derivation of the PMPMWFS signal obtained from the reflected light of the composite optical resonator.

\subsection{Model for the PMPMWFS derivation.}
\label{SUBSEC:cal_model}
Two EOMs are placed in series immediately after the laser source. The laser beam is phase-modulated by the EOMs to generate two PM sidebands $f_{\Const{a}}$ and $f_{\Const{b}}$ ($f_{\Const{a}}>f_{\Const{b}}$), and is then injected into the composite optical resonator. The light reflected from the composite optical resonator is detected by a QPD. The QPD signal is demodulated at the difference frequency $f_{\Const{a}}-f_{\Const{b}}$ between the two PM sideband frequencies. After the demodulated signal passes through a low-pass filter, the PMPMWFS signal is obtained. \par
When the angles of the mirrors constituting the composite optical resonator fluctuate, the reflectance of the first-order Gaussian beam components change for the carrier and the two PM sidebands. Therefore, by describing the PMPMWFS signal in terms of these reflectance, we derive how the mirror misalignment is manifested in the PMPMWFS signal.

\subsection{Derivation of the electric field reflected from the composite optical cavity.}

\label{SUBSEC:refl_electricfield} 

The time variation of the output electric field from the laser ($E_{\Const{1}}$) can be expressed as follows:
\begin{equation}
\label{EQ:E1}
E_{\Const{1}}=E_{\Const{0}}e^{i \Omega \Var{t}},
\end{equation}
where $E_{\Const{0}}$ and $\Omega$ are the amplitude and angular frequency of the laser, respectively.\par
The electric field $E_{\Const{inc}}$, when the sidebands $f_{\Const{a}}$ and $f_{\Const{b}}$ are phase-modulated in series, can be expressed as follows:
\begin{equation}
\label{EQ:Einc}
\frac{E_{\Const{inc}}}{E_{\Const{1}}}=e^{ i({m_{\Const{a}}\cos{\omega_{\Const{a}}}\Var{t}+m_{\Const{b}}\cos{\omega_{\Const{b}}}\Var{t}})},
\end{equation}
where the modulation angular frequencies are denoted by $\Const \omega_{\Const{a}}$ and $\omega_{\Const{b}}$, and the modulation indices are denoted by $m_{\Const{a}}$ and $m_{\Const{b}}$. \par
Expanding Eq. \ref{EQ:Einc} into a series using the Bessel function $J_{n}(m_{\Const{a}})$ and $J_{n}(m_{\Const{b}})$ gives the following expression:
\begin{equation}
\frac{E_{\Const{inc}}}{E_{\Const{1}}}=(\sum_{n=-\infty}^\infty J_{n}(m_{\Const{a}})(i)^{n}e^{in\Const \omega_{\Const{a}}\Var{t}})(\sum_{n=-\infty}^\infty J_{n}(m_{\Const{b}})(i)^{n}e^{in\omega_{\Const{b}}\Var{t}}).
\end{equation}
Assuming both modulation indices satisfy $m_{\Const{a}},m_{\Const{b}}\ll 1$, the higher-order terms of the Bessel function are negligible. Therefore, $E_{\Const{inc}}$ can be approximated by retaining the Bessel terms up to the second order, yielding the following expression:
\begin{equation}
   \label{EQ:Einc_J}
   \begin{split}
      \frac{E_{\Const{inc}}}{E_{\Const{1}}}&\simeq J_{0}(m_{\Const{a}})J_{0}(m_{\Const{b}})
      +iJ_{1}(m_{\Const{a}})J_{0}(m_{\Const{b}})(e^{i\Const \omega_{\Const{a}}\Var{t}}+e^{-i\Const \omega_{\Const{a}}\Var{t}})+iJ_{0}(m_{\Const{a}})J_{1}(m_{\Const{b}})(e^{i\omega_{\Const{b}}\Var{t}}+e^{-i\omega_{\Const{b}}\Var{t}})\\
      &-J_{1}(m_{\Const{a}})J_{1}(m_{\Const{b}})(e^{i(\omega_{\Const{a}}+\omega_{\Const{b}})\Var{t}}+e^{-i(\Const \omega_{\Const{a}}+\omega_{\Const{b}})\Var{t}})-J_{1}(m_{\Const{a}})J_{1}(m_{\Const{b}})(e^{i(\Const \omega_{\Const{a}}-\omega_{\Const{b}})\Var{t}}+e^{-i(\omega_{\Const{a}}-\omega_{\Const{b}})\Var{t}}).\\
      &-J_{2}(m_{\Const{a}})J_{0}(m_{\Const{b}})(e^{2i\Const \omega_{\Const{a}}\Var{t}}+e^{-2i\Const \omega_{\Const{a}}\Var{t}})-J_{0}(m_{\Const{a}})J_{2}(m_{\Const{b}})(e^{2i\Const \omega_{\Const{b}}\Var{t}}+e^{-2i\Const \omega_{\Const{b}}\Var{t}}).
   \end{split}
\end{equation}

In the next section, the reflected light from the interferometer is detected by a QPD and demodulated at the difference frequency $f_{\Const{a}}-f_{\Const{b}}$. Therefore, only the $f_{\Const{a}}-f_{\Const{b}}$ component is discussed there. The $f_{\Const{a}}-f_{\Const{b}}$ component of the QPD signal consists of the beat component between the carrier and the difference frequency, the beat component between the $f_{\Const{a}}$ and $f_{\Const{b}}$ sidebands, and the beat component between the second-harmonic and sum frequency. However, the beat components at the second harmonic and sum frequencies scale with the fourth power of the modulation index andthus become negligibly small compared to the other beat components, which scale with the second power.
Consequently, the second harmonic and sum frequency components $E_{\Const{inc}}$ in Eq. \ref{EQ:Einc_J} can be neglected and are omitted hereafter.\par
Therefore, light phase-modulated with two modulation frequencies $f_{\Const{a}}$ and $f_{\Const{b}}$ produces carrier light, the associated PM sidebands, and difference frequency components. \par
As stated in Ref. \cite{Hefetz:97}, the reflection response of the interferometer can be expressed in matrix form after extending Eq. \ref{EQ:Einc_J} to include spatial modes.
The reflection matrices for the carrier light and the modulated sidebands (upper: u, lower: l) are distinguished by subscripts as $\bm{M}_{\Const{c}}$, $\bm{M}_{\Const{ua}}$, $\bm{M}_{\Const{la}}$, $\bm{M}_{\Const{ub}}$, $\bm{M}_{\Const{lb}}$, $\bm{M}_{\Const{u(a-b)}}$, and $\bm{M}_{\Const{l(a-b)}}$.
Using these matrices, the reflected light from the interferometer ($E_{\Const{ref}}$) can be expressed as follows:

\begin{equation}
   \label{EQ:Eref}
   \begin{split}
   &\frac{E_{\Const{ref}}}{E_{\Const{1}}}=
   \left[
   \begin{array}{ll}
       u^{\Const{00}} & u^{\Const{10}} 
   \end{array}
   \right]
   \Biggl(
    \bm{M}_{\Const{c}}
    \left[
    \begin{array}{ll}
       J_{0}(m_{\Const{a}})J_{0}(m_{\Const{b}}) \\\noalign{\vskip3pt}
       0
    \end{array}
    \right]\\
   &+\bm{M}_{\Const{ua}}
   \left[
   \begin{array}{ll}
   iJ_{1}(m_{\Const{a}})J_{0}(m_{\Const{b}}) \\\noalign{\vskip3pt}
   0
   \end{array}
   \right] e^{i\omega_{\Const{a}}t}
   +\bm{M}_{\Const{la}}
   \left[
   \begin{array}{ll}
   iJ_{1}(m_{\Const{a}})J_{0}(m_{\Const{b}}) \\\noalign{\vskip3pt}
   0
   \end{array}
   \right] e^{-i\omega_{\Const{a}}t}\\
   &+\bm{M}_{\Const{ub}}
   \left[
   \begin{array}{ll}
   iJ_{0}(m_{\Const{a}})J_{1}(m_{\Const{b}})\\\noalign{\vskip3pt}
   0
   \end{array}
   \right]e^{i\omega_{\Const{b}}t}
   +\bm{M}_{\Const{lb}}
   \left[
   \begin{array}{ll}
   iJ_{0}(m_{\Const{a}})J_{1}(m_{\Const{b}})\\\noalign{\vskip3pt}
   0
   \end{array}
   \right]e^{-i\omega_{\Const{b}}t}\\
   &+\bm{M}_{\Const{u(a-b)}}
   \left[
   \begin{array}{ll}
      -J_{1}(m_{\Const{a}})J_{1}(m_{\Const{b}}) \\\noalign{\vskip3pt}
      0
   \end{array}
   \right] e^{i(\omega_{\Const{a}}-\omega_{\Const{b}})t}
   +\bm{M}_{\Const{l(a-b)}}
   \left[
   \begin{array}{ll}
      -J_{1}(m_{\Const{a}})J_{1}(m_{\Const{b}}) \\\noalign{\vskip3pt}
      0
   \end{array}
   \right] e^{-i(\omega_{\Const{a}}-\omega_{\Const{b}})t}
   \Biggr),
\end{split}
\end{equation}
we assume the optical axis misalignment is small and the reflected light is dominated by the TEM00 and TEM10 modes, so that the electric field amplitude can be expressed using the TEM00 ($u^{\Const{00}}$) and TEM10 ($u^{\Const{10}}$) \cite{Hefetz:97}.

\subsection{Derivation of PMPMWFS signals.}
   
\label{SUBSEC:demosignal}
The intensity $P_{\Const{qpd}}$ measured by a QPD can be expressed as follows in terms of the reflected electric field $E_{\Const{qpd}}$:
\begin{equation}
   P_{\Const{qpd}}=E_{\Const{qpd}}E_{\Const{qpd}}^{\ast}.
\end{equation}
$E_{\Const{qpd}}$ represents the electric field at the QPD that detects the \mbox{interferometer's} reflected light, and is calculated by incorporating the light propagation matrix from the interferometer to the QPD into Eq. \ref{EQ:Eref}. This propagation matrix is characterized by the distance $l_{\Const{pd}}$ from the \mbox{interferometer's} reflection point to the QPD and the Gouy phase $\eta_{\Const{pd}}$ associated with the light propagation between them \cite{Hefetz:97}.

\begin{equation}\label{EQ:Eref2Pref}
   \begin{split}
      \frac{E_{\Const{qpd}}}{E_{\Const{1}}}&=e^{-ikl_{\Const{pd}}}\{E_{\Const{c}}+E_{\Const{ua}}e^{i\omega_{\Const{a}}(t-\frac{l_{\Const{pd}}}{\Const{c}})}+E_{\Const{la}}e^{-i\omega_{\Const{a}}(t-\frac{l_{\Const{pd}}}{\Const{c}})}+E_{\Const{ub}}e^{i\omega_{\Const{b}}(t-\frac{l_{\Const{pd}}}{\Const{c}})}+E_{\Const{lb}}e^{-i\omega_{\Const{b}}(t-\frac{l_{\Const{pd}}}{\Const{c}})}\\
&+E_{\Const{u(a-b)}}e^{i(\omega_{\Const{a}}-\omega_{\Const{b}})(t-\frac{l_{\Const{pd}}}{\Const{c}})}+E_{\Const{l(a-b)}}e^{-i(\omega_{\Const{a}}-\omega_{\Const{b}})(t-\frac{l_{\Const{pd}}}{\Const{c}})}\},
\end{split}
\end{equation}
\begin{equation}
   \label{EQ:Eref_AtoE}
   \begin{split}
   E_{\Const{c}} &=
   \left[
   \begin{array}{ll}
      u^{\Const{00}} & u^{\Const{10}} 
   \end{array}
   \right]
   \left[
   \begin{array}{ll}
       e^{i\eta_{\Const{pd}}} & 0 \\\noalign{\vskip3pt}
        0 & e^{2i\eta_{\Const{pd}}}
   \end{array}
   \right]
   \bm{M}_{\Const{c}}
   \left[
   \begin{array}{ll}
      J_{0}(m_{\Const{a}})J_{0}(m_{\Const{b}}) \\\noalign{\vskip3pt}
      0
   \end{array}
   \right], 
   \\[6pt]
   E_{x\Const{a}} &=
   \left[
   \begin{array}{ll}
      u^{\Const{00}} & u^{\Const{10}} 
   \end{array}
   \right]
   \left[
   \begin{array}{ll}
       e^{i\eta_{\Const{pd}}} & 0 \\\noalign{\vskip3pt}
        0 & e^{2i\eta_{\Const{pd}}}
   \end{array}
   \right]
   \bm{M}_{x\Const{a}}
   \left[
   \begin{array}{ll}
      iJ_{0}(m_{\Const{b}})J_{1}(m_{\Const{a}}) \\\noalign{\vskip3pt}
      0
   \end{array}
   \right], 
   \\[6pt]
   E_{x\Const{b}} &=
   \left[
   \begin{array}{ll}
      u^{\Const{00}} & u^{\Const{10}} 
   \end{array}
   \right]
   \left[
   \begin{array}{ll}
   e^{i\eta_{\Const{pd}}} & 0 \\\noalign{\vskip3pt}
       0 & e^{2i\eta_{\Const{pd}}}
   \end{array}
   \right]
   \bm{M}_{x\Const{b}} 
   \left[
   \begin{array}{ll}
      iJ_{0}(m_{\Const{a}})J_{1}(m_{\Const{b}})\\\noalign{\vskip3pt}
      0
   \end{array}
   \right], 
   \\[6pt]
   E_{x\Const{(a-b)}} &=
   \left[
   \begin{array}{ll}
      u^{\Const{00}} & u^{\Const{10}} 
   \end{array}
   \right]
   \left[
   \begin{array}{ll}
       e^{i\eta_{\Const{pd}}} & 0 \\\noalign{\vskip3pt}
       0 & e^{2i\eta_{\Const{pd}}}
   \end{array}
   \right] 
   \bm{M}_{x\Const{(a-b)}}
   \left[
   \begin{array}{ll}
      -J_{1}(m_{\Const{a}})J_{1}(m_{\Const{b}}) \\\noalign{\vskip3pt}
      0
   \end{array}
   \right], \quad x=(\Const{u},\Const{l}).\nonumber
   \end{split}
   \end{equation}
To demodulate $P_{\Const{qpd}}$ at two sideband difference frequencies, the angular frequency $\omega_{\Const{a}}-\omega_{\Const{b}}$ components in $P_{\Const{qpd}}$ are summarized as follows:
\begin{equation}\label{EQ:P_qpd}
   \begin{split}
      \frac{\{P_{\Const{qpd}}\}_{\Const{\omega_{a}-\Const{\omega_{b}}}}}{|E_{\Const{1}}|^2}
      &=2Re\{(E_{\Const{c}}E_{\Const{l(a-b)}}^{\ast}+E_{\Const{c}}^{\ast}E_{\Const{u(a-b)}}+E_{\Const{ua}}E_{\Const{ub}}^{\ast}+E_{\Const{la}}^{\ast}E_{\Const{lb}})e^{i(\omega_{\Const{a}}-\omega_{\Const{b}})(t-\frac{l_{\Const{pd}}}{\Const{c}})}\}.
   \end{split}
\end{equation}

When the horizontal, vertical, and light propagation directions are taken as the x-, y-, and z-directions, respectively, the amplitude distributions of $u^{\Const{00}}$ and $u^{\Const{10}}$ on the x-z plane are expanded under the assumption tilting along the x-axis \cite{siegman1986lasers}:
\begin{equation}\label{EQ:U0}
   u^{\Const{00}}(x,z)=\biggl(\frac{2}{\pi\omega^2(z)}\biggr)^{\frac{1}{4}}\exp\biggl[-\biggl(\frac{x}{\omega(z)}\biggr)^2-i\frac{kx^{2}}{2R(z)}\biggr],
\end{equation}
\begin{equation}\label{EQ:U1}
   u^{\Const{10}}(x,z)=\biggl(\frac{2}{\pi\omega^2(z)}\biggr)^{\frac{1}{4}}\frac{2x}{\omega(z)}\exp\biggl[-\biggl(\frac{x}{\omega(z)}\biggr)^2-i\frac{kx^{2}}{2R(z)}\biggr],
\end{equation}
where the curvature radius of the beam wavefront is denoted by $R(z)$, and the beam radius at position $z$ is denoted by $\omega(z)$. As shown in Eq. \ref{EQ:U0} and Eq. \ref{EQ:U1}, $u^{\Const{00}}(x,z)$ is an even function of $x$, whereas $u^{\Const{10}}(x,z)$ is an odd function of $x$. Consequently, the terms ${\lvert u^{\Const{00}}\rvert}^2$ and ${\lvert u^{\Const{10}}\rvert}^2$ are both even, while the terms $u^{\Const{00}}u^{\Const{10}\ast}$ and $u^{\Const{10}}u^{\Const{00}\ast}$ are odd. 
The WFS signal is obtained by taking the difference between the positive and negative regions along the x-axis using a photo diode divided into two halves along the x-axis like a QPD. Consequently, the terms ${\lvert u^{\Const{00}}\rvert}^2$ and ${\lvert u^{\Const{10}}\rvert}^2$ are canceled out, while only the cross terms $u^{\Const{00}}u^{\Const{10}\ast}$ and $u^{\Const{10}}u^{\Const{00}\ast}$ in Eq. \ref{EQ:P_qpd} contribute to the signal.
\begin{equation}
   \label{EQ:U0U1}
   2\int_{0}^{\infty}u^{\Const{00}}(x,z)u^{\Const{10}\ast}(x,z)dx=2\int_{0}^{\infty}u^{\Const{00}\ast}(x,z)u^{\Const{10}}(x,z)dx=\sqrt{\frac{2}{\pi}}.
\end{equation}

Given that Eq. \ref{EQ:U0U1} and $J_{0}(m_{\Const{a}})J_{0}(m_{\Const{b}})J_{1}(m_{\Const{a}})J_{1}(m_{\Const{b}})$ are real-valued, the PMPMWFS signal($P_{\Const{PMPMWFS}}$) is obtained as follows:
\begin{equation}\label{EQ:PMPMWFS}
   \begin{split}
      &\frac{P_{\Const{PMPMWFS}}}{2\sqrt{\frac{2}{\pi}}J_{0}(m_{\Const{a}})J_{0}(m_{\Const{b}})J_{1}(m_{\Const{a}})J_{1}(m_{\Const{b}})|E_{\Const{1}}|^2}\\
      &=Re[\{(-M_{\Const{c}}^{\Const{00}\ast}M_{\Const{u(a-b)}}^{\Const{10}}-M_{\Const{c}}^{\Const{10}}M_{\Const{l(a-b)}}^{\Const{00}\ast}+M_{\Const{ua}}^{\Const{10}} M_{\Const{ub}}^{\Const{00}\ast} +  M_{\Const{la}}^{\Const{00}\ast}  M_{\Const{lb}}^{\Const{10}})e^{i\eta_{\Const{pd}}}\\
      &+(-M_{\Const{c}}^{\Const{00}}M_{\Const{l(a-b)}}^{\Const{10}\ast}-M_{\Const{c}}^{\Const{10}\ast}M_{\Const{u(a-b)}}^{\Const{00}}+M_{\Const{ua}}^{\Const{00}}  M_{\Const{ub}}^{\Const{10}\ast} +  M_{\Const{la}}^{\Const{10}\ast}  M_{\Const{lb}}^{\Const{00}})e^{-i\eta_{\Const{pd}}}\}e^{i(\omega_{\Const{a}}-\omega_{\Const{b}})(t-\frac{l_{\Const{pd}}}{\Const{c}})}].
   \end{split}
\end{equation}
We distinguish the $00$ and $10$ components of the reflection matrix $\bm{M}$ for the composite optical resonator by superscripts such as $M^{\Const{00}}$ and $M^{\Const{10}}$. Furthermore, under the condition that the misalignment is sufficiently small, $M^{\Const{10}}$ can be expressed linearly with respect to the angle of each mirror 
\cite{Hefetz:97}. A conventional WFS detects the misalignment signal between the carrier and the PM sideband. From Eq. \ref{EQ:PMPMWFS}, it is evident that the PMPMWFS signal detects the relative misalignment between the sideband $f_{\Const{a}}$ and $f_{\Const{b}}$, as well as between the carrier and the difference frequency $f_{\Const{a}}-f_{\Const{b}}$. It was also demonstrated that the PMPMWFS signal varies due to the Gouy phase from the interferometer reflection to a QPD.

\section{PMPMWFS signal in the PRXARM of KAGRA.}
\label{sec:PRXARM}
To verify the response of the PMPMWFS signal, we performed theoretical calculations and experimental measurements using the PRXARM at KAGRA. The detailed configuration of the PRXARM employed in this Sec. \ref{sec:PRXARM} is outlined in Sec. \ref{sec:config_PRXARM}. 
In Section \ref{sec:PRXARM_cal}, we derive the PMPMWFS signals for angular variations of each mirror in the PRXARM configuration presented in Sec. \ref{sec:config_PRXARM}, and compare with the conventional WFS signal using the carrier and PM sideband beat signal. In Section \ref{sec:measurement_cal_PRXARM}, we experimentally measure the signals in \mbox{KAGRA's} PRXARM.
In Section \ref{sec:feedback}, we demonstrate feedback control using the PMPMWFS signal on \mbox{KAGRA's} PRXARM.\par

\subsection{PRXARM of KAGRA}
\label{sec:config_PRXARM}
\begin{figure}[!h]
\centering
\includegraphics[width=140mm]{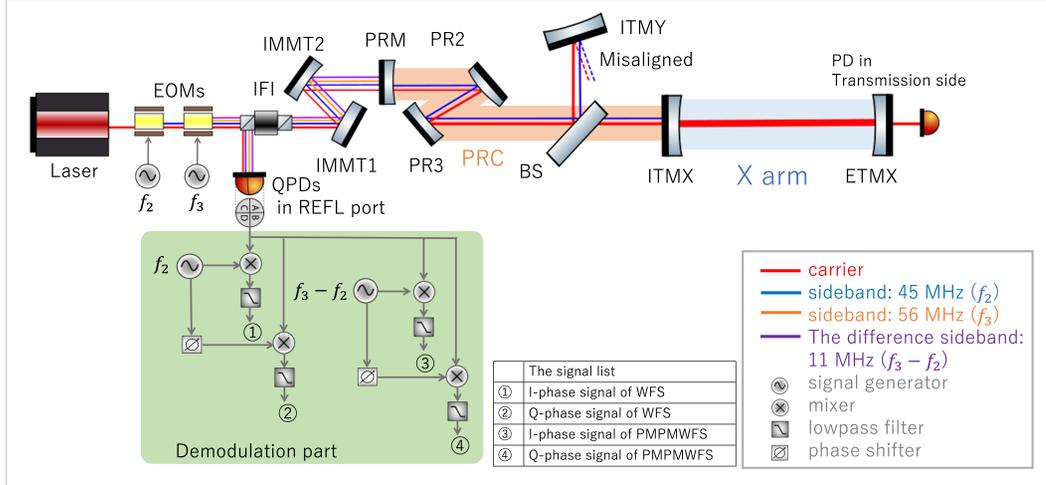}
\caption{Schematic of the PRXARM configuration in KAGRA. Red, blue, orange, and purple lines indicate the optical paths for the carrier, $f_\Const{2}$ sideband, $f_\Const{3}$ sideband, and difference frequency $f_\Const{3}-f_\Const{2}$, respectively. The WFS and PMPMWFS signals are obtained by demodulating the QPD signals on the REFL port at frequencies $f_\Const{2}$ and $f_\Const{3}-f_\Const{2}$, respectively, and then passing them through low-pass filters. The demodulated WFS signal is separated into the I-phase signal (at the reference demodulation phase) and the Q-phase signal (phase-shifted by 90 degrees from the reference). The X-arm cavity transmission power is detected by a photodiode at its transmission port. }
\label{FIG:KAGRA_PRXARM}
\end{figure}

The PRXARM in KAGRA is an optical resonator consisting of the X-arm cavity and PRC (Fig. \ref{FIG:KAGRA_PRXARM}). The mirrors constituting the PRXARM are summarized below. The X-arm cavity consists of Input Test Mass X (ITMX) and End Test Mass X (ETMX) mirrors. PRC comprises PRM, Power Recycling mirror 2 (PR2), 3 (PR3) and ITMX. The convex mirror PR2 and the concave mirror PR3 are placed to adjust the Gouy phase between PRM and ITMX. Input Mode-Matching Telescope 1 and 2 (IMMT1 and IMMT2) are mirrors used to adjust the incident optical axis into PRC. For the PRXARM measurement, to avoid interference from the Y-arm cavity, ITMY was misaligned.\par
The incident light power was 1 W. 
In KAGRA, the Input Mode Cleaner (IMC) is installed to remove higher-order spatial modes present in the light incident on the interferometer. The \mbox{IMC's} free spectral range (FSR = 5.63 MHz) is used as the seed frequency, and the frequencies of the phase-modulated sidebands are set to integer multiples of this seed frequency ($f_\Const{seed}$). Specifically, the light is phase-modulated at the 8th harmonic ($f_\Const{2}$ = 45.02 MHz) and the 10th harmonic ($f_\Const{3}$ = 56.27 MHz) of $f_\Const{seed}$. When the carrier resonates in the entire PRXARM, the optical sideband field at $f_\Const{2}$ resonates in PRC and anti-resonates with the X-arm cavity, while the optical sideband fields at $f_\Const{3}$ and at the difference frequency $f_\Const{3}-f_\Const{2}$ anti-resonate with the entire composite optical resonator.\par

\subsection{Theoretical calculations for the PRXARM.}

\label{sec:PRXARM_cal}

This section discusses the PMPMWFS signal obtained in the PRXARM. \par
We derive the reflected signal response when the PRM, PR2, PR3, ITMX, and ETMX mirrors are each tilted by 0.1 nrad, in order to calculate the I-phase and Q-phase signals of the WFS signal using Eq. \ref{EQ:PMPMWFS} in PRXARM. The signal responses are shown in Tab. \ref{tb:mirror_response_carrierf2}, \ref{tb:mirror_response_f3} and \ref{tb:mirror_response_diff}.\par

\begin{table}[htbp]
   \centering
   \begin{tabular}{|l|ccc|}
   \hline
   mirror & $M_{\Const{c}}^{\Const{10}}$ & $M_{\Const{u}(f_\Const{2})}^{\Const{10}}$ & $M_{\Const{l}(f_\Const{2})}^{\Const{10}}$ \\
   \hline
   
     PR3  & $-5.6\times10^{2} + 4.5\times10^{2}i$
     & $-6.3\times10^{4} - 1.1\times10^{4}i$
     & $-1.9\times10^{4} + 6.2\times10^{4}i$ \\

PR2  & $-1.4\times10^{2} + 23i$
     & $-7.7\times10^{3} - 1.3\times10^{3}i$
     & $-2.3\times10^{3} + 7.5\times10^{3}i$ \\

PRM  & $7.9\times10^{3} - 2.3\times10^{4}i$
     & $-2.4\times10^{4} + 1.2\times10^{4}i$
     & $8.3\times10^{3} + 2.5\times10^{4}i$ \\

ITMX & $-8.6\times10^{3} - 4.4\times10^{3}i$
     & $3.0\times10^{4} + 5.2\times10^{3}i$
     & $9.1\times10^{3} - 3.0\times10^{4}i$ \\

ETMX & $1.5\times10^{4} + 6.9\times10^{3}i$
     & $-2.4\times10^{2} - 42 i$
     & $56 - 1.8\times10^{2}i$ \\

   \hline
   \end{tabular}
\caption{Each element of reflection matrices for the carrier and $f_\Const{2}$. Each value represents the Matrix element normalized by the inclination of each mirror.}
\label{tb:mirror_response_carrierf2}
\end{table}

\begin{table}[htbp]
   \centering
   \begin{tabular}{|l|cc|}
      \hline
      mirror & $M_{\Const{u}(f_\Const{3})}^{\Const{10}}$ & $M_{\Const{l}(f_\Const{3})}^{\Const{10}}$ \\
      \hline
      
      PR3  & $-3.5\times10^{4} - 4.2\times10^{4}i$
     & $2.3\times10^{4} + 1.8\times10^{4}i$ \\

PR2  & $-4.2\times10^{3} - 5.1\times10^{3}i$
     & $2.9\times10^{3} + 2.2\times10^{3}i$ \\

PRM  & $9.5\times10^{3} - 2.3\times10^{4}i$
     & $5.7\times10^{3} - 2.4\times10^{4}i$ \\

ITMX & $1.7\times10^{4} + 2.0\times10^{4}i$
     & $-1.1\times10^{4} - 8.8\times10^{3}i$ \\

ETMX & $31 + 37 i$
     & $57 + 44 i$ \\


      \hline
      \end{tabular}
   \caption{Each element of reflection matrices for $f_\Const{3}$. Each value represents the Matrix element normalized by the inclination of each mirror.}
   \label{tb:mirror_response_f3}
   \end{table}

\begin{table}[htbp]
      \centering
      \begin{tabular}{|l|cc|}
         \hline
   mirror & $M_{\Const{u}(f_\Const{3}-f_\Const{2})}^{\Const{10}}$ & $M_{\Const{l}(f_\Const{3}-f_\Const{2})}^{\Const{10}}$ \\
   \hline
   PR3  & $-4.8\times10^{3} - 9.8\times10^{3}i$
     & $1.3\times10^{4} - 2.6\times10^{3}i$ \\

PR2  & $-5.8\times10^{2} - 1.2\times10^{3}i$
     & $1.6\times10^{3} - 3.2\times10^{2}i$ \\

PRM  & $7.0\times10^{3} + 2.5\times10^{4}i$
     & $-2.1\times10^{4} + 1.5\times10^{4}i$ \\

ITMX & $2.3\times10^{3} + 4.7\times10^{3}i$
     & $-6.3\times10^{3} + 1.3\times10^{3}i$ \\

ETMX & $3.4 + 6.9 i$
     & $46 - 9.3 i$ \\

   \hline
      \end{tabular}
   \caption{Each element of reflection matrices for $f_\Const{3}-f_\Const{2}$. Each value represents the Matrix element normalized by the inclination of each mirror.}
   \label{tb:mirror_response_diff}
   \end{table}

As described in Sec. \ref{SUBSEC:demosignal}, the PMPMWFS signal comprises the component arising from the beat signal between the $f_\Const{2}$ and $f_\Const{3}$ sidebands, as well as from the beat between the carrier and the difference frequency. Consequently, the results are presented together. The conventional WFS signal \cite{Morrison:94} was demodulated at the REFL port using $f_\Const{2}$. To compare it with the PMPMWFS calculation results, it was calculated under identical conditions as in Sec. \ref{SEC:PMPMWFS_CAL} using the following WFS equation ($P_{\Const{WFS}}$):
\begin{equation}\label{EQ:P_WFS}
   \begin{split}
   &\frac{P_{\Const{WFS}}}{2\sqrt{\frac{2}{\pi}}J_{0}(m_{\Const{a}})J_{0}^{2}(m_{\Const{b}})J_{1}(m_{\Const{a}})|E_{\Const{1}}|^2}\nonumber\\
   &=Re[i\{(-M_{\Const{c}}^{10}M_{\Const{la}}^{00\ast}+M_{\Const{c}}^{00\ast}M_{\Const{ua}}^{10})e^{i\eta_{\Const{pd}}}+(-M_{\Const{c}}^{00}M_{\Const{la}}^{10\ast}+M_{\Const{c}}^{10\ast}M_{\Const{ua}}^{00})e^{-i\eta_{\Const{pd}}}\}e^{i\omega_{\Const{a}}(t-\frac{l_{\Const{pd}}}{c})}].
   \end{split}
\end{equation}

\begin{figure}[!h]
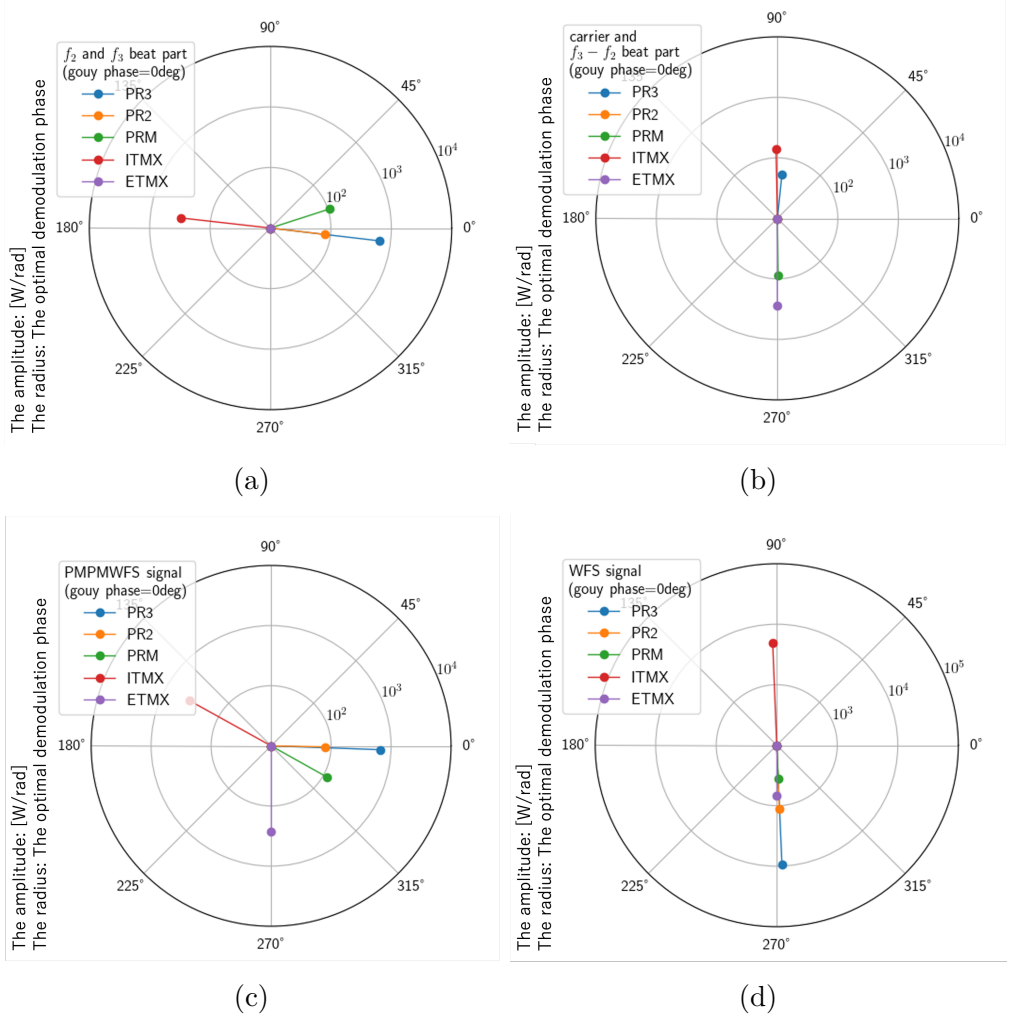

   \centering
   \begin{subfigure}{0.40\textwidth}
       \centering
       \includegraphics[width=\linewidth]{./fig/cal_result/PRXARM_cal_gouy0_f2f3_refETMX.pdf}
       \caption{}
       \label{FIG:cal_PRXARM_a}
   \end{subfigure}
   \begin{subfigure}{0.40\textwidth}
       \centering
       \includegraphics[width=\linewidth]{./fig/cal_result/PRXARM_cal_gouy0_carrierf2f3_refETMX.pdf}
       \caption{}
       \label{FIG:cal_PRXARM_b}
   \end{subfigure}\\
   \begin{subfigure}{0.40\textwidth}
       \centering
       \includegraphics[width=\linewidth]{./fig/cal_result/PRXARM_cal_gouy0_all_PMPMWFS_refETMX.pdf}
       \caption{}
       \label{FIG:cal_PRXARM_c}
   \end{subfigure}
   \begin{subfigure}{0.40\textwidth}
       \centering
       \includegraphics[width=\linewidth]{./fig/cal_result/PRXARM_cal_gouy0_carrierf2_refETMX.pdf}
       \caption{}
       \label{FIG:cal_PRXARM_d}
   \end{subfigure}
   \caption{I-Q plane plot calculated for the PMPMWFS signal and conventional WFS signal in PRXARM. The radial axis is plotted on a logarithmic scale. Figure \ref{FIG:cal_PRXARM_a} shows the beat component between the $f_\Const{2}$ and $f_\Const{3}$ sidebands. Figure \ref{FIG:cal_PRXARM_b} shows the beat component between the carrier and the difference frequency $f_\Const{3}-f_\Const{2}$ sideband, and Fig. \ref{FIG:cal_PRXARM_c} shows the PMPMWFS signal obtained by combining these components, while Fig. \ref{FIG:cal_PRXARM_d} shows the calculated WFS signal. The argument represents the optimal demodulation phase in degrees at which the signal from each mirror is maximized. The magnitude represents the signal normalized by the tilt of each mirror. Note that the Gouy phase was calculated assuming 0 degrees. Furthermore, the demodulation phase for all degrees of freedom was rotated so that the ETMX angle signal in the I-phase component was minimized for both PMPMWFS and conventional WFS results. The rotated phase applied in Fig. \ref{FIG:cal_PRXARM_c} was also applied in Fig. \ref{FIG:cal_PRXARM_a} and \ref{FIG:cal_PRXARM_b}.}
   \label{FIG:cal_PRXARM}
\end{figure}

At first, we plotted the result in polar coordinates on the I-Q plane, where the horizontal and vertical axes of the I-Q plane represent the amplitudes of the I-phase and Q-phase signals, respectively. The I-phase and Q-phase signals denote the QPD signals demodulated by a demodulation frequency that is set at a certain phase and the phase 90 degrees shifted from it, respectively.
The demodulation phase adjusted to minimize the ETMX angle signal in the I-phase component (Fig. \ref{FIG:cal_PRXARM}). Note that the Gouy phase was set to 0 degrees in these calculations.\par

\begin{figure}[!h]
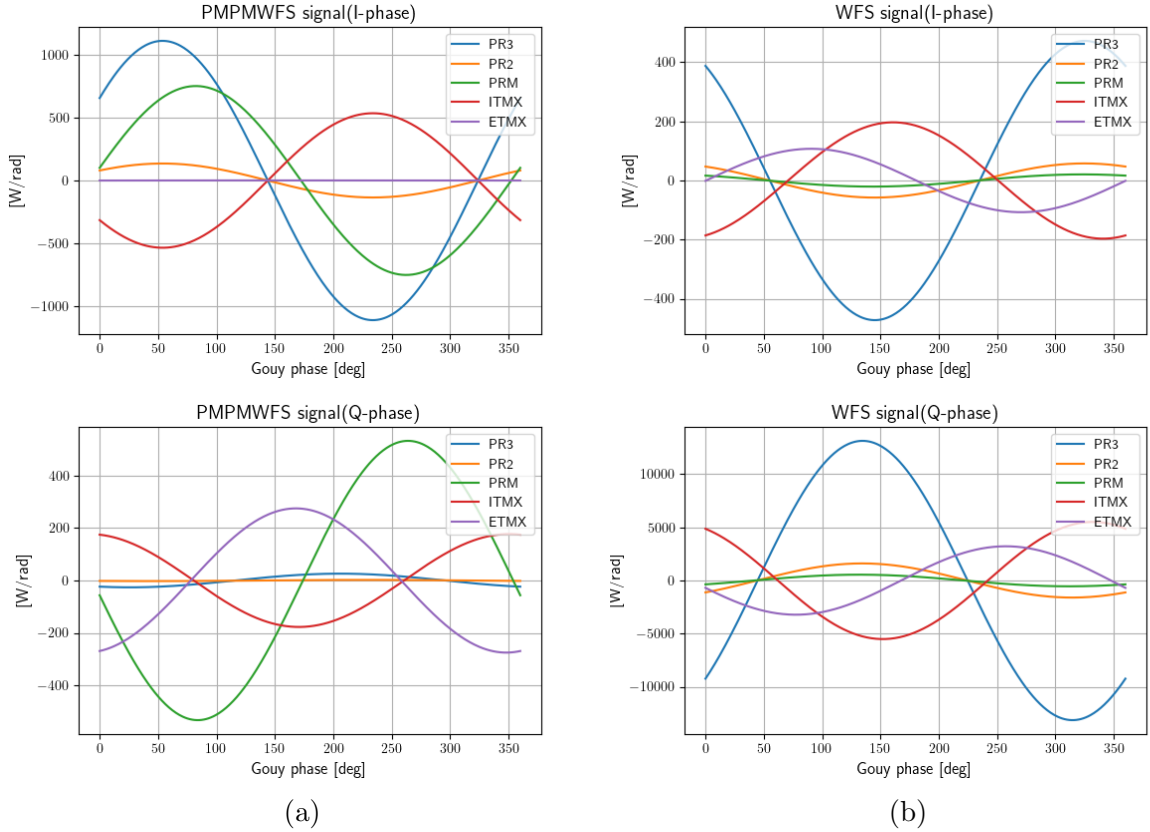

   \centering
   \begin{subfigure}{0.48\textwidth}
       \centering
       \includegraphics[width=\linewidth]{./fig/cal_result/gouyphase/PRXARM_cal_gouysweep_Iphase_refETMX_all_PMPMWFS.pdf}\\
       \includegraphics[width=\linewidth]{./fig/cal_result/gouyphase/PRXARM_cal_gouysweep_Qphase_refETMX_all_PMPMWFS.pdf}\\
       \caption{}
       \label{FIG:cal_PRXARM_gouy_a}
   \end{subfigure}
   \begin{subfigure}{0.48\textwidth}
       \centering
       \includegraphics[width=\linewidth]{./fig/cal_result/gouyphase/PRXARM_cal_gouysweep_Iphase_refETMX_carrierf2.pdf}\\
       \includegraphics[width=\linewidth]{./fig/cal_result/gouyphase/PRXARM_cal_gouysweep_Qphase_refETMX_carrierf2.pdf}\\
       \caption{}
       \label{FIG:cal_PRXARM_gouy_b}
   \end{subfigure}
   \caption{Gouy phase dependence of the magnitude of I-phase and Q-phase signals when the Gouy phase $\eta_{\Const{pd}}$ is swept from -180 to 180 degrees. The horizontal axis represents the Gouy phase $\eta_{\Const{pd}}$ degrees, and the vertical axis represents the signal magnitude normalized by the mirror angle. The Figures \ref{FIG:cal_PRXARM_gouy_a} and \ref{FIG:cal_PRXARM_gouy_b} show the graphs of the PMPMWFS and WFS signals, respectively. The upper and lower figures show the graphs of I-phase and Q-phase signals, respectively. The demodulation phase is set such that the ETMX angle signal is minimized in the I-phase at the Gouy phase of 0 degrees.}
   \label{FIG:cal_PRXARM_gouy}
\end{figure}

Figure \ref{FIG:cal_PRXARM_gouy} shows the Gouy phase dependence of the magnitude of the I-phase and Q-phase signals. The calculation in Fig. \ref{FIG:cal_PRXARM_gouy} was performed using the demodulation phase determined in Fig. \ref{FIG:cal_PRXARM}. The results reveal the following.\par

\begin{itemize}
   \item The amplitude of the PMPMWFS signal is generally less than one-tenth of that of the WFS signal. This is because WFS uses the beat between the one PM sideband and the carrier, whereas PMPMWFS is demodulated based on the beat component between the two PM sidebands. In this calculation, the modulation indices for sideband 2 and sideband 3 are set to 0.15. As a result, the amplitude is reduced due to differences in the coefficients of the Bessel functions.\par
   \item The ETMX angle signal is scarcely visible in the beat signal between $f_\Const{2}$ and $f_\Const{3}$. This is likely because that both $f_\Const{2}$ and $f_\Const{3}$ are anti-resonant with the X-arm cavity, making them insensitive to the misalignment of the X-arm \mbox{cavity's} optical axis.\par
   \item By choosing an appropriate Gouy phase, the PRM angle signal can be detected with higher sensitivity than those of the other mirrors.\par
   \item The optimal demodulation phase for the ETMX angle signal and the other mirrors is close to 90 degrees in the PMPMWFS signal. Conversely, in the conventional WFS signal, the phase difference between each mirror angle signal is close to 0 degrees or 180 degrees. Consequently, the PMPMWFS signal enables efficient decoupling of the ETMX angle signal from the others.\par
\end{itemize}


\subsection{Comparison between the measurement and the theoretical calculation in the PRXARM}
\label{sec:measurement_cal_PRXARM}
An optical system was constructed using \mbox{KAGRA's} PRXARM to acquire both the PMPMWFS and the conventional WFS signals. In particular, QPDs were positioned to achieve the Gouy phases of -55 degrees and 50 degrees, respectively, in order to separate the degrees of freedom for translational and tilt misalignments along the relative optical axis. 
Each mirror was oscillated at a frequency of 1.3 Hz, separately in the pitch and yaw directions. The PMPMWFS and WFS signals were measured and normalized by the tilt of each mirror, which was monitored using the optical lever.

\begin{figure}[!h]
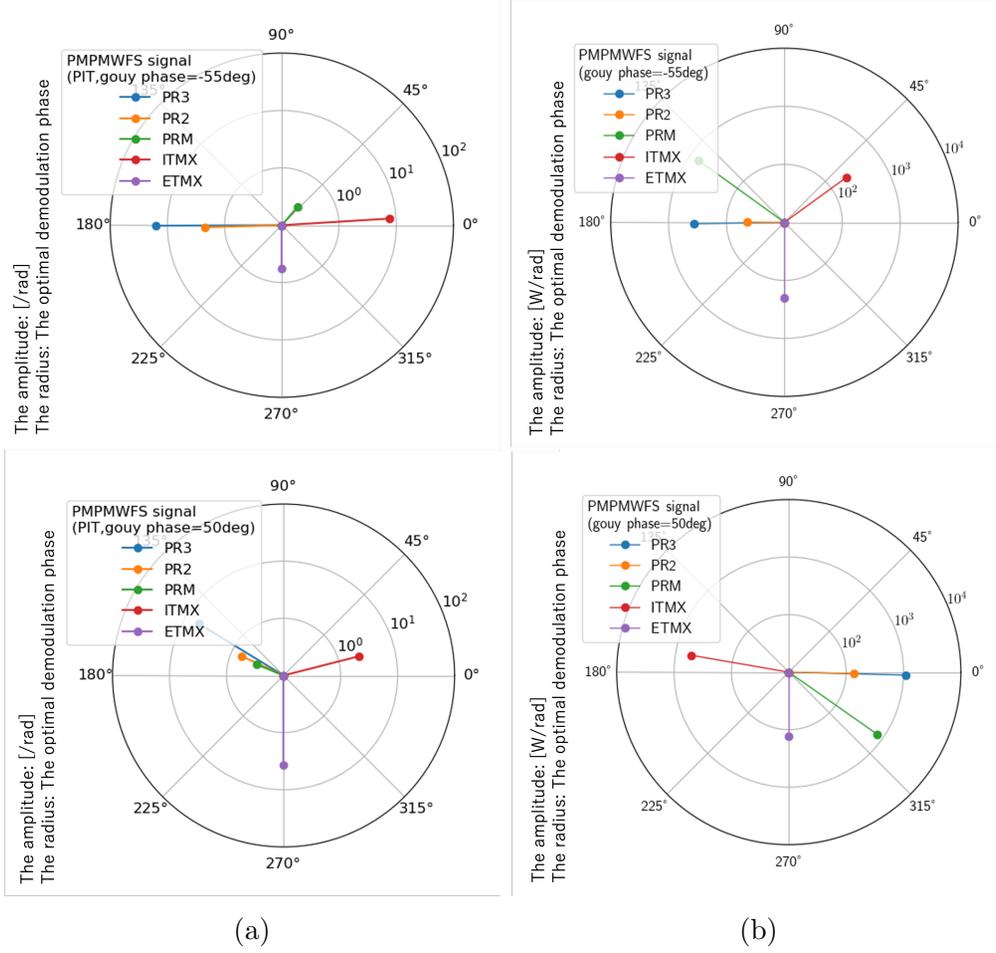

   \centering
   \begin{subfigure}{0.40\textwidth}
       \centering
       \includegraphics[width=\linewidth]{./fig/measurement_result/demo_refETMX/Sensmat_PRXARM_PIT_refETMX_-55_PMPMWFS.pdf}\\
       \includegraphics[width=\linewidth]{./fig/measurement_result/demo_refETMX/Sensmat_PRXARM_PIT_refETMX_50_PMPMWFS.pdf}
       \caption{}
       \label{FIG:measure_PRXARM_a}
   \end{subfigure}
   \begin{subfigure}{0.40\textwidth}
       \centering
       \includegraphics[width=\linewidth]{./fig/cal_result/PRXARM_cal_gouy-55_all_PMPMWFS_refETMX.pdf}\\
       \includegraphics[width=\linewidth]{./fig/cal_result/PRXARM_cal_gouy50_all_PMPMWFS_refETMX.pdf}
       \caption{}
       \label{FIG:measure_PRXARM_c}
   \end{subfigure}
   \caption{Measured results and theoretical calculation results for the PRXARM. Figure \ref{FIG:measure_PRXARM_a} shows the PMPMWFS signals obtained by tilting each mirror in the pitch angle. The signals in Fig. \ref{FIG:measure_PRXARM_a} were obtained by normalizing the measured values to the local tilt of each mirror measured by the optical lever, when each mirror was periodically tilted at 1.3 Hz. Figure \ref{FIG:measure_PRXARM_c} presents the corresponding theoretical results. The upper row of each plot corresponds to the Gouy phase of -55 degrees, while the lower row corresponds to the Gouy phase of 50 degrees. Each plot shows the signals with the demodulation phase rotated to minimize the ETMX angle signal in the I-phase component.}
   \label{FIG:PRXARM_result_PIT_PMPMWFS}
\end{figure}

The results, plotted on the I-Q plane based on measurements taken with each mirror tilted in the pitch angle, are shown in Figure \ref{FIG:measure_PRXARM_a} and \ref{FIG:measure_PRXARM_b}. To compare with the calculations, the I-Q plane plots of the PMPMWFS and WFS signals calculated at the Gouy phases of -55 degrees and 50 degrees are shown in Fig. \ref{FIG:measure_PRXARM_c} and \ref{FIG:measure_PRXARM_d}, respectively. Note that the demodulation phase here is rotated such that the ETMX angle signal reached its minimum at I-phase. From these graphs, the following can be observed:\par

\begin{figure}[!h]
   \centering
   \begin{subfigure}{0.40\textwidth}
      \centering
      \includegraphics[width=\linewidth]{./fig/measurement_result/demo_refETMX/Sensmat_PRXARM_PIT_refETMX_-55_WFS.pdf}\\
      \includegraphics[width=\linewidth]{./fig/measurement_result/demo_refETMX/Sensmat_PRXARM_PIT_refETMX_50_WFS.pdf}
      \caption{}
      \label{FIG:measure_PRXARM_b}
  \end{subfigure}
   \begin{subfigure}{0.40\textwidth}
       \centering
       \includegraphics[width=\linewidth]{./fig/cal_result/PRXARM_cal_gouy-55_carrierf2_refETMX.pdf}\\
       \includegraphics[width=\linewidth]{./fig/cal_result/PRXARM_cal_gouy50_carrierf2_refETMX.pdf}
       \caption{}
       \label{FIG:measure_PRXARM_d}
   \end{subfigure}
   \caption{Measured results and theoretical calculation results for the PRXARM. Figure \ref{FIG:measure_PRXARM_b} shows the WFS signals obtained by tilting each mirror in the pitch angle.  The signals in Fig. \ref{FIG:measure_PRXARM_b} were obtained by normalizing the measured values to the local tilt of each mirror measured by the optical lever, when each mirror was periodically tilted at 1.3 Hz. Figure \ref{FIG:measure_PRXARM_d} presents the corresponding theoretical results. The upper row of each plot corresponds to the Gouy phase of -55 degrees, while the lower row corresponds to the Gouy phase of 50 degrees. Each plot shows the signals with the demodulation phase rotated to minimize the ETMX angle signal in the I-phase component.}
   \label{FIG:PRXARM_result_PIT_WFS}
\end{figure}

\begin{itemize}
   \item As predicted by theoretical calculations, the amplitude of the PMPMWFS signal is consistently less than one-tenth of that of the conventional WFS signal.\par
   \item It is either in phase or opposite phase with all degree of freedom in the conventional WFS signals, whereas the optimal demodulation phase for the PRM and ETMX angle signals in the PMPMWFS signals is orthogonal to that of the other angle signals.\par
   \item Comparison of the experimental and theoretical results for the PMPMWFS signal, the optimal demodulation phase of the PRM angle signal is different. This difference may be caused by the larger misalignment from the optimal position during the experiment, although the exact cause remains unclear.
   \item The results for some degrees of freedom exhibit sign flips. In particular, the PMPMWFS results at a Gouy phase of 50 degrees in Fig. \ref{FIG:measure_PRXARM_a} show a sign flip in all degrees of freedom except for the ETMX angle signal, whereas the WFS results at a Gouy phase of 50 degrees in Fig. \ref{FIG:measure_PRXARM_b} show a sign flip only for the ITMX angle signal. These sign flips can be explained by the Gouy phases of 79 degrees and 63 degrees in the PMPMWFS and WFS calculations, respectively. However, it is unlikely that the Gouy phase at the QPD location differs by more than 12 degrees, considering the beam waist position and Rayleigh range estimated from the measured beam profile. The cause of these discrepancies remains unclear.
\end{itemize}
Based on the above results, the PMPMWFS has an advantage in decoupling the arm-axis fluctuation from the others compared with the conventional WFS. However, it should be noted that the signal amplitude is reduced due to the Bessel-function terms.\par
\begin{figure}[!h]
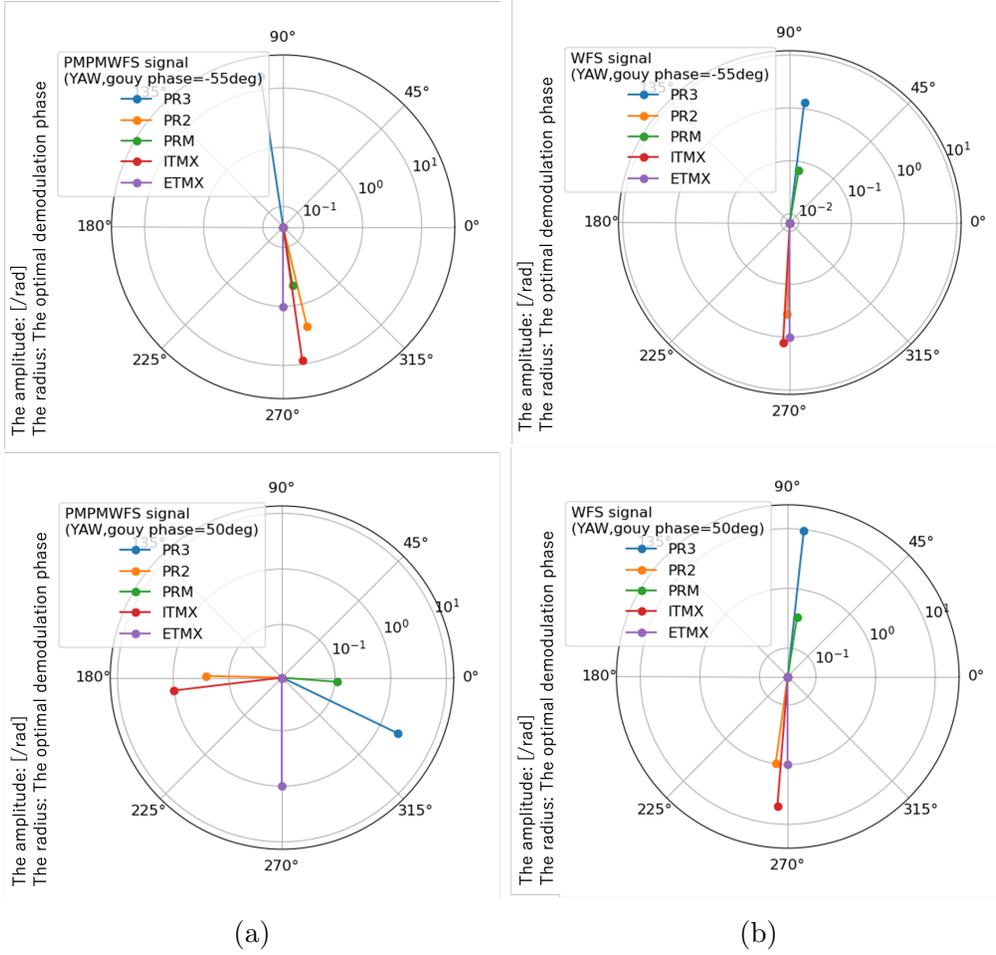

   \centering
   \begin{subfigure}{0.40\textwidth}
      \centering
      \includegraphics[width=\linewidth]{./fig/measurement_result/demo_refETMX/Sensmat_PRXARM_YAW_refETMX_-55_PMPMWFS.pdf}\\
      \includegraphics[width=\linewidth]{./fig/measurement_result/demo_refETMX/Sensmat_PRXARM_YAW_refETMX_50_PMPMWFS.pdf}
      \caption{}
      \label{FIG:measure_PRXARM_PMPMWFS_YAW}
  \end{subfigure}
   \begin{subfigure}{0.40\textwidth}
       \centering
       \includegraphics[width=\linewidth]{./fig/measurement_result/demo_refETMX/Sensmat_PRXARM_YAW_refETMX_-55_WFS.pdf}\\
       \includegraphics[width=\linewidth]{./fig/measurement_result/demo_refETMX/Sensmat_PRXARM_YAW_refETMX_50_WFS.pdf}
       \caption{}
       \label{FIG:measure_PRXARM_WFS_YAW}
   \end{subfigure}
   \caption{Measurement results for the PRXARM. Figure \ref{FIG:measure_PRXARM_PMPMWFS_YAW} and \ref{FIG:measure_PRXARM_WFS_YAW} show the PMPMWFS and WFS signals obtained by tilting each mirror in the yaw angle, respectively. The signals in Fig. \ref{FIG:measure_PRXARM_a} and \ref{FIG:measure_PRXARM_b} were obtained by normalizing the angular motions of each mirror measured by the optical lever, when each mirror was periodically tilted at 1.3 Hz. The upper row of each plot corresponds to the Gouy phase of -55 degrees, while the lower row corresponds to the Gouy phase of 50 degrees. Each plot shows the signals with the demodulation phase rotated to minimize the ETMX angle signal in the I-phase component.}
   \label{FIG:PRXARM_result_YAW}
\end{figure}

Measurements in the yaw angle were also shown in Fig. \ref{FIG:PRXARM_result_YAW}. The results similarly showed that the PMPMWFS signal detects ETMX orthogonally. The Section \ref{sec:feedback} examines whether control can actually be implemented based on these results.\par

\clearpage

\subsection{Demonstration of alignment control with PMPMWFS signals}
\label{sec:feedback}
To verify the controllability of the optical axis using the PMPMWFS signal, the alignment sensing and control using PMPMWFS was implemented in \mbox{KAGRA's} PRXARM, based on the results presented in Sec. \ref{sec:measurement_cal_PRXARM}.\par
Figure \ref{FIG:measure_PRXARM_c} and \ref{FIG:measure_PRXARM_d} show that the pitch and yaw angle signals of PR3 were significantly detected by the QPD1 signal at the Gouy phase of -55 degrees. Therefore, the PMPMWFS signal detected by the QPD1 signal was used as the PR3 control. Furthermore, since the optical axis in the PRXARM consists of three axes, namely the X-arm cavity, the PRC, and the incident-light axis, three mirrors were necessary to correct the overall relative misalignment. Consequently, the Alignment Dither System (ADS) \cite{Kawabe1994} was implemented for the mirrors that were not controlled by the PMPMWFS feedback signal. 
The demodulation phase of the QPD1 signal was adjusted so that the magnitude of the PR3 pitch signal was maximized in the I-phase signal, and the demodulation phase of the signals of the QPD at Gouy phase of 50 degrees (QPD2) was adjusted so that the ETMX I-phase signal was minimized. Furthermore, the ETMX and PRM angle signals were decoupled by appropriately combining two QPD signals. Consequently, the pitch signal ($S_{\Const{pitch}}$) and yaw signal ($S_{\Const{yaw}}$) for PMPMWFS, obtained from the I-phase signal ($S_{\Const{X}}^{\Const{I}}$) and Q-phase signal ($S_{\Const{X}}^{\Const{Q}}$) from QPDX, are expressed as follows.\par
\begin{equation}
   \label{EQ:feedback_pit}
   S_{\Const{pitch}}=2.232S_{\Const{1}}^{\Const{I}}+1.000S_{\Const{2}}^{\Const{Q}}
\end{equation}
\begin{equation}
   \label{EQ:feedback_yaw}
   S_{\Const{yaw}}=2.599S_{\Const{1}}^{\Const{I}}-9.522S_{\Const{2}}^{\Const{I}}+1.000S_{\Const{2}}^{\Const{Q}}
\end{equation}
At this time, $S_{\Const{pitch}}$ and $S_{\Const{yaw}}$ obtained when each mirror was tilted by 1 µrad are summarized in Tab. \ref{tb:PRXARM_errorsignal_dof}. The signal amplitudes of the ETMX and PRM angle signal are less than one-tenth of those of the PR3 angle signal, indicating that the signals are well decoupled.\par

\begin{figure}[!h]
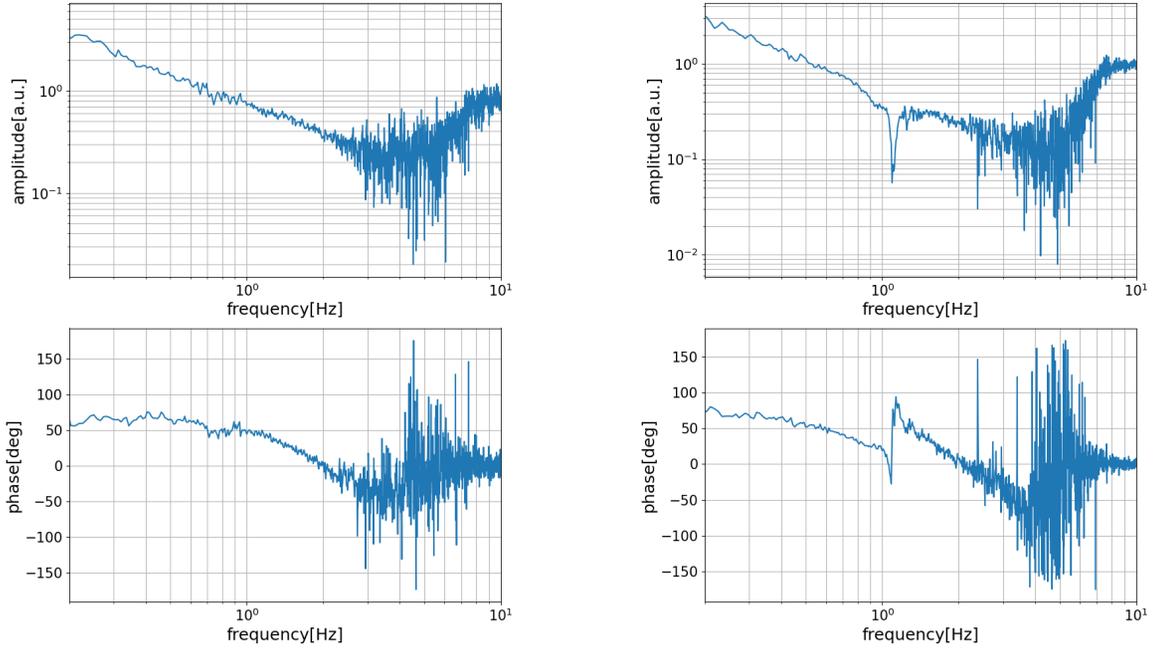

    \begin{tabular}{cc}
        \begin{minipage}{.5\textwidth}
            \centering
            \includegraphics[width=70mm]{fig/measurement_result/OLTF_PIT_PRXARM.pdf}
            \label{fig_first}
        \end{minipage}
        \begin{minipage}{.5\textwidth}
            \centering
            \includegraphics[width=70mm]{fig/measurement_result/OLTF_YAW_PRXARM.pdf}
            
            \label{fig_second}
        \end{minipage}
    \end{tabular}
    \caption{OLTF of PR3 control with $S_{\Const{pitch}}$ and $S_{\Const{yaw}}$. The left figure is OLTF of the pitch control. The right figure is OLTF of the yaw control. Since the signal-to-noise ratio above 2 Hz is low, we show the data only up to 10 Hz.}
    \label{FIG:OLTF}
\end{figure}

\begin{table}[!h]
  \centering
  \begin{tabular}{|l|ccccc|}
  \hline
[Count/urad] & ITMX & ETMX & PRM & PR2 & PR3\\\hline
Error signal of pitch & 22.77 & 3.988 & 0.3700 & 5.837 & 53.59 \\
Error signal of yaw   & 74.73 & 4.554 & 3.730 & 20.72 & 136.2 \\
\hline
  \end{tabular}
  \caption{$S_{\Const{pitch}}$ and $S_{\Const{yaw}}$ values when each mirror was tilted by 1urad.}
  \label{tb:PRXARM_errorsignal_dof}
\end{table}
Pitch and yaw of PR3 were controlled by $S_{\Const{pitch}}$ and $S_{\Const{yaw}}$.
Figure \ref{FIG:OLTF} shows the Open Loop Transfer Functions (OLTFs) of the pitch and yaw controls. The Unity Gain Frequencies (UGFs) for pitch and yaw were 0.4 Hz and 0.6 Hz, respectively, and the phase margin at the UGFs was 60 degrees and 50 degrees, respectively, indicating robust control performance.\par

\begin{figure}[!h]
\centering
\includegraphics[width=140mm]{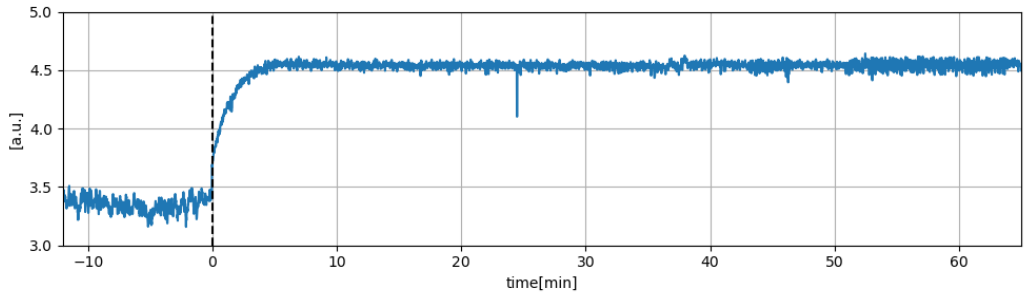}
\caption{Transmission power of the X-arm cavity, normalized by the transmission power when the X-arm cavity is resonating with misaligned PRM. The black dotted line indicates when all ASC turned on. }
\label{FIG:long_time_series_PRXARM}
\end{figure}

With the WFS control enabled, the interferometer kept locking stably for more than 1 hour as shown in Fig. \ref{FIG:long_time_series_PRXARM}. When the control was turned on, the transmission power increased. This observation confirms that the ASC using PMPMWFS is effective for interferometer alignment sensing and control.\par

\clearpage
\section{Conclusion.}
To decouple signals from the arm cavity axis misalignment to the other optical axis misalignments, we propose a new WFS technique (PMPMWFS) that demodulates the difference frequency using two sidebands that do not resonate in the arm cavities. We derived the theoretical equations and compared them with the experiments using the PRXARM in KAGRA. The PMPMWFS signal measures the relative misalignment between the carrier and the difference frequency $f_{\Const{a}}-f_{\Const{b}}$, and the relative misalignment between two sidebands, thereby demonstrating a signal response that is distinct from that of the WFS signal. Furthermore, in the PRXARM, signal separation was achieved because the ETMX angle signals appeared orthogonally. Finally, we demonstrated alignment control using the PMPMWFS signal, thereby validating this technique as a viable detection signal. \par

\section*{Acknowledgment}
This work was supported by JST SPRING, Grant Number JPMJSP2121, 
Ministry of Education, Culture, Sports, Science and Technology (MEXT), Japan Society for the Promotion of Science (JSPS) Leading-edge Research Infrastructure Program, JSPS Grant-in-Aid for Specially Promoted Research 26000005, JSPS Grant-in-Aid for Scientific Research on Innovative Areas 2402: 24103006, 24103005, and 2905: JP17H06358, JP17H06361 and JP17H06364, JSPS Core-to-Core Program A. Advanced Research Networks, JSPS Grant-in-Aid for Scientific Research (S) 17H06133 and 20H05639 , JSPS Grant-in-Aid for Transformative Research Areas (A) 20A203: JP20H05854, the joint research program of the Institute for Cosmic Ray Research, University of Tokyo, National Research Foundation (NRF), Computing Infrastructure Project of Global Science experimental Data hub Center (GSDC) at KISTI, Korea Astronomy and Space Science Institute (KASI), and Ministry of Science and ICT (MSIT) in Korea, Academia Sinica (AS), AS Grid Center (ASGC) and the National Science and Technology Council (NSTC) in Taiwan under grants including the Science Vanguard Research Program, Advanced Technology Center (ATC) of NAOJ, and Mechanical Engineering Center of KEK.

\clearpage
\section*{Appendix.Parameters used in the PRXARM calculation}
The parameters used in the calculations for Sec. \ref{sec:PRXARM_cal} and Sec. \ref{sec:measurement_cal_PRXARM} are summarised in Tab. \ref{tb:para}.
\begin{table}[htbp]
\raggedright
\begin{tabular}{|l|cc|}
\hline
 & power [W] & wavelength [nm] \\\hline
laser&1.0 & 1064.0 \\  \hline
\end{tabular}
\begin{tabular}{|l|ccc|}
\hline
EOM & frequency [MHz] & modulation index & modulation type  \\\hline
& $f_{\Const{seed}}$=5.627 & &   \\
EOM2& $f_{\Const{seed}}$*8.0 & 0.15 & phase modulation \\ 
EOM3& $f_{\Const{seed}}$*10.0 &0.15 & phase modulation \\ 
\hline
\end{tabular}
\begin{tabular}{|l|cc|}
\hline
space & length [m] & gouy phase [deg]  \\\hline
from IMMT1 to IMMT2 & 3.105 & 1.965  \\ 
from IMMT2 to PRM & 4.8416 & 4.519 \\ 
from PRM to PR2 & 14.7615 & 14.124  \\ 
from PR2 to PR3 & 11.0661 & 1.278 \\ 
from PR3 to BS & 15.7638 & 0.225 \\ 
from BS to ITMX & 26.6649 &0.391 \\ 
from ITMX to ETMX & 3.000e3 & 126.078 \\ \hline
\end{tabular}
\begin{tabular}{|l|cccc|}
\hline
mirror & Transmittance & Loss & curvature radius [m] & incident angle [deg]\\
\hline
IMMT1 & 0.0078 & 0.00 & -8.91 & 3.6\\
IMMT2 & 0.0078 & 0.00 & 14.005 & 3.6\\
PRM & 0.10 & 45.0e-6 & 458.1285 & \\
PR2 & 500.0e-6 & 45.0e-6 & -3.0764 & 0.6860 \\
PR3 & 50.0e-6 & 45.0e-6 & -24.9165 & 0.6860 \\
BS & 0.50 & 0.00  &  $\infty$ & 45.0\\ 
ITMX & 0.0040 & 45.0e-6 & 1900.0 & \\
ETMX & 5.0e-06 & 45.0e-6 & 1900.0 & \\
\hline
\end{tabular}
\caption{Parameters used in the theoretical calculations. The radius of curvature of each mirror is defined as positive for the concave side with respect to its non-black surface, as shown in Fig. \ref{FIG:KAGRA_PRXARM}. For ITMX, the mirror thickness (0.15 m) and the refractive index of the substrate ($n_{\Const{ITMX}}=1.754$) were also taken into account in the calculations.}
\label{tb:para}
\end{table}

\let\doi\relax


\end{document}